\documentclass[aps,twocolumn,prl,showpacs]{revtex4}
\usepackage{amsmath}
\usepackage{amssymb}
\usepackage{epsfig}
\usepackage{colordvi}

\def\cchi{{\boldsymbol{\chi}}}
\def\ee{\boldsymbol{\cal{E}}}
\def\FF{{\bf F}}
\def\KK{{\bf K}}
\def\pp{{\bf P}}
\def\llambda{{\boldsymbol{\lambda}}}
\def\ppsi{{\boldsymbol{\Psi}}}
\def\pt{{\bf P}_{\rm t}}
\def\XX{{\bf X}}
\def\ZZ{{\bf Z}}

\begin{document}


\title{
First-Principles Calculations at Constant Polarization
}

\author{Oswaldo Di\'eguez}
\author{David Vanderbilt}

\affiliation{
Department of Physics and Astronomy, Rutgers University, 
Piscataway, New Jersey 08854-8019, USA
}

\begin{abstract}
We develop an exact formalism for performing first-principles calculations for 
insulators at fixed electric polarization.
As shown by Sai, Rabe, and Vanderbilt (SRV) [N.~Sai, K.~M.~Rabe, and
D.~Vanderbilt, Phys.~Rev.~B~{\bf 66}, 104108 (2002)], who designed an
approximate method to tackle the same problem,
such an approach allows one to map out the energy landscape as a function
of polarization, providing a powerful tool for the theoretical investigation
of polar materials.
We apply our method to a system in which the ionic contribution to the
polarization dominates (a broken-inversion-symmetry perovskite),
one in which this is not the case (a III-V semiconductor), and one in
which an additional degree of freedom plays an important role
(a ferroelectric phase of KNO$_3$).  We find that while the SRV method
gives rather accurate results in the first case, the present approach
provides important improvements to the physical description in the
latter cases.
\end{abstract}

\date{\today}


\maketitle


\marginparwidth 2.7in
\marginparsep 0.5in

\def\odm#1{\marginpar{\Blue{\small OD: #1}}}
\def\dvm#1{\marginpar{\Red{\small DV: #1}}}



In 1993 King-Smith and Vanderbilt \cite{KingSmith1993PRB} introduced a theory
for computing the electric polarization of an infinite solid, showing for the 
first time that this property was indeed a well-defined quantity for an
insulating material.
Their theory of the bulk polarization (TBP) is based on computing the 
electronic contribution to the polarization as a Berry phase of the
valence-band Bloch wave functions transported across the Brillouin zone.
A practical numerical scheme to compute it was given in the same 
paper \cite{KingSmith1993PRB}, and it is now widely used in first-principles
calculations.
Later, Souza, \'I\~niguez, and Vanderbilt \cite{Souza2002PRL}
(SIV) used the TBP as the cornerstone for a method to calculate the 
exact ground state of an insulator in the presence of an
electric field by minimizing an electric enthalpy functional expressed
in terms of occupied Bloch-like states on a uniform grid of points in
reciprocal space.
It is therefore possible to compute from first-principles many interesting 
properties of materials related to their behavior under electric fields.

In this Letter we propose a method to do first-principles calculations not at
constant electric field, but at constant electric polarization.
In this way, it would become possible to map the energy $E$ of an insulator as
a function of its polarization ${\bf P}$.
There are several reasons why this is useful.
First, it allows for a exhaustive search for competing local minima and
for saddle points in a system with a complicated energy surface.
These features would otherwise be hidden in a standard optimization
procedure, which would only be likely to find a single miminum of the energy.
Second, knowing $E({\bf P})$, we can calculate various properties related to
derivatives of $E$ with respect to ${\bf P}$, such as the linear and
non-linear dielectric susceptibilities.  While methods of
density-functional perturbation theory (DFPT)
\cite{deGironcoli1989PRL_Giannozzi1991PRB_Gonze1997PRB} can also be used
to access these susceptibilities, our approach is advantageous in that
it does not require much special-purpose programming and yields
non-linear susceptibilities with very little extra effort beyond that
needed to obtain the linear ones.
And third, Landau-Devonshire theories have historically provided
an important avenue to the understanding of ferroelectric
materials, based on an expansion of the energy in powers of
polarization.  The ability to compute $E(\bf P)$ may open the door
to the first-principles derivation of Landau-Devonshire descriptions,
as opposed to the usual empirical formulations, for a wide range of
ferroelectric materials.

The problem of performing simulations at constant polarization in the
context of first-principles calculations has been addressed previously.
In particular, Sai, Rabe, and Vanderbilt \cite{Sai2002PRB} (SRV) developed
an {\em approximate} method to do this and applied it to obtain $E(\bf
P)$ for several perovskite systems of interest. 
A similar approach was used by Fu and Bellaiche to study the electromechanical
response of solids under finite electric fields \cite{Fu2003PRL}.
Our method relies on many of the ideas of the SRV one, but it is instead
an {\em exact} method because it incorporates the SIV theory of
finite electric fields \cite{Souza2002PRL}.

\begin{figure}[b]
\center{\epsfig{file=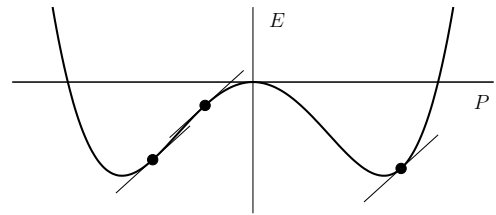,width=2.6in}}
\caption{Sketch of energy as a function of polarization for a 
         double-well potential.
         The value of the electric field needed to impose the polarization
         is the same at the three indicated points, as their slopes are
         the same.}
\label{fig_dw}
\end{figure}

Our goal is to find the minimum energy $E$ of a system of atoms for which
the electric polarization takes some target value $\pt$.
The system has both ionic and electronic degrees of freedom, and we
will use vectors $\XX$ and $\ppsi$ to represent them.
Vector $\XX$ has $3N+6$ components given by the cartesian coordinates of the 
$N$ atoms in the simulation cell and the 6 components of the strain tensor.
On the other hand, $\ppsi$ contains the one-electron wave functions of the
system.
We are therefore facing a standard constrained optimization problem that can 
be solved by introducing a Lagrange multiplier $\llambda$ and searching for
the minimum of $E+\llambda \cdot (\pp - \pt)$, or, discarding the constant 
term, the minimum of  $E+\llambda \cdot \pp$. 
Defining $\llambda = - \Omega \ee$, where $\Omega$ is the volume of the unit
cell of the material under consideration, we have
\begin{eqnarray}
&& \min_{\genfrac{}{}{0pt}{}{\XX,\ppsi}{\llambda \rightarrow \pp=\pt}} 
 \{ E(\XX,\ppsi) 
+ 
\llambda \cdot \pp(\XX,\ppsi) \}
\nonumber \\
&& \quad \quad =
\min_{\genfrac{}{}{0pt}{}{\XX}{\ee \rightarrow \pp=\pt}}
\{
\min_{\ppsi}
\{ E(\XX,\ppsi) - \Omega \ee \cdot \pp(\XX,\ppsi) \}
\}
\nonumber \\
&& \quad \quad = 
\min_{\genfrac{}{}{0pt}{}{\XX}{\ee \rightarrow \pp=\pt}} 
H(\XX,\ee),
\end{eqnarray}
where $\llambda \rightarrow \pp=\pt$ indicates a minimization over the values
of $\llambda$ that impose the constraint $\pp=\pt$.
The function $H(\XX,\ee)$ is the electric enthalpy that is minimized in the 
SIV method to find the electronic ground-state of an insulator in the presence 
of an electric field $\ee$ \cite{Souza2002PRL}.

Therefore, the problem of finding the minimum energy of a system that is 
constrained to have polarization $\pp$ is equivalent to finding the ground 
state of a system under an electric field $\ee$ that imposes polarization
$\pp$.
However, the apparently straightforward approach of doing
calculations at different values of the electric
field to find $E(\pp)$ parametrically will not work in general. 
Fig.~\ref{fig_dw} illustrates why this is so for a typical double-well 
potential characteristic of ferroelectric perovskites.
In this case, several values of the polarization correspond to the {\it same}
value of the electric field, since $\ee = \Omega^{-1}(dE/d\pp)$, and an 
optimization using the SIV theory will
most likely find the point that has the lowest energy of the three, i.e.,
the global minimum of $H(P)$.
With extra care it might be possible to find the secondary local minimum
of $H(P)$, but not the third solution, which is a maximum (or, in 3D, a
saddle point) of $H(P)$.
It follows from this line of reasoning that one cannot map
the region of $E(P)$ that has negative curvature,
and it may also prove difficult to map nearby regions
of positive curvature.

We now describe how to perform a minimization of the enthalpy $H$ over the
ionic degrees of freedom $\XX$ and the electric fields $\ee$ that produce a
desired polarization $\pt$ in a way that parallels the presentation in
Sec.~III.A of Ref.~\onlinecite{Sai2002PRB}.
From now on, we will assume that the optimization is done while keeping the
cell vectors fixed.  (It is possible to remove this constraint, but it is
necessary to take into account some technical subtleties related to the
way in which the stress tensor is computed in the presence of an electric 
field.  The details, together with examples,
will be presented elsewhere \cite{WuIP}.)
We begin with a trial guess $(\XX_0,\ee_0)$, and we expand $H(\XX,\ee)$ 
and $\pp(\XX,\ee)$ as low-order Taylor series
\begin{eqnarray}
H &=& H_0 
      - \FF \, \delta \XX 
      + \frac{1}{2} \, \delta \XX \, \KK \, \delta \XX
      - \Omega \pp \, \delta \ee , \\
\pp &=& \pp_0 
        + \frac{1}{\Omega} \ZZ \, \delta \XX 
        + \frac{1}{4 \pi} \cchi \, \delta \ee ,
\end{eqnarray}
where $H_0 = H(\XX_0,\ee_0)$ and $\pp_0 = \pp(\XX_0,\ee_0)$
(atomic units are used throughout unless stated otherwise).
The $F_i = - (\partial H/\partial X_i)$ are the $3N$ components
of the forces on the atoms at finite electric field \cite{Souza2002PRL},
the $K_{ij} = (\partial^2 H/\partial X_i \partial X_j)$ are the
$3N \times 3N$ force-constant matrix elements, the
$Z_{i \alpha} = \Omega (\partial P_{\alpha}/\partial X_i)$
are the $3N \times 3$ Born effective charges, and the $\chi_{\alpha\beta}
= 4 \pi (\partial P_{\beta}/\partial {\cal E}_{\alpha})$ are the 
$3 \times 3$ elements of the dielectric susceptibility tensor.
We can then predict that $\partial H/\partial X_i=0$ and $\pp=\pt$
at an $\XX=\XX_0+\delta\XX$ and an $\ee=\ee_0+\delta\ee$ given by
\begin{equation}
\left( \begin{array}{cc}
  \KK                    &  -\ZZ  \\
  -\ZZ^{T}  &  -\frac{\Omega}{4\pi}\,\cchi
\end{array} \right)
\left( \begin{array}{c}
  \delta \XX  \\
  \delta \ee
\end{array} \right)
=
\left( \begin{array}{c}
  \FF  \\
  \Omega\Delta\pp_0
\end{array} \right) ,
\label{eq_system}
\end{equation}
where $\Delta \pp_0=\pp_0-\pt$.
This system of linear equations can be solved iteratively, refining 
$\delta \XX$ and $\delta \ee$ until $\FF$ and $\Delta\pp_0$ both vanish.
As mentioned before, $\FF$ can be computed exactly in the presence
of an electric field according to the SIV prescription \cite{Souza2002PRL}.
The guiding tensors $\KK$, $\ZZ$ and $\cchi$ do not need to be computed
exactly; replacing these by approximate versions only results in a slower
convergence to the correct solution, without shifting the solution
itself.  In particular, we normally find it sufficient to compute
$\KK$, $\ZZ$ and $\cchi$ at zero electric field.

The scheme described here has been implemented to work with the
ABINIT \cite{abinit} density-functional theory (DFT) \cite{Hohenberg1964PR}
code.
Instead of performing a direct iterative solution of Eq.~(\ref{eq_system}),
we have used a nested-loop algorithm for the sake of robustness.
Starting with some guess for $\XX$ and $\ee$,
we keep the atoms fixed and vary the field in the internal
loop until the polarization is the target one, a problem that is well 
behaved.
Once this is achieved, we solve Eq.~(\ref{eq_system}) to get new values of
$\XX$ and $\ee$, and iterate until convergence is achieved.

\begin{figure}
\center{\epsfig{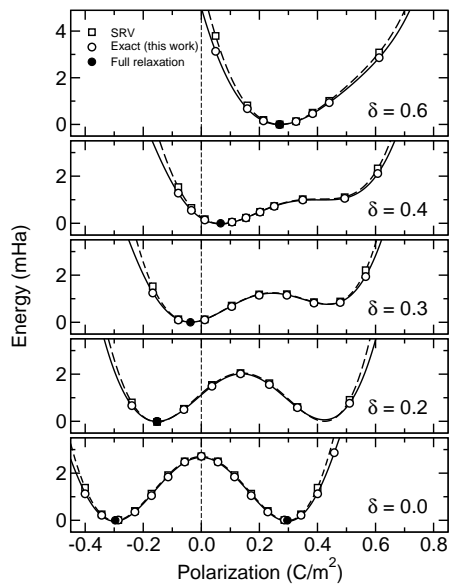}}
\caption{Energy versus polarization along the $c$ axis for the
broken-inversion-symmetry perovskite described in the text, for different
values of $\delta$.
The solid (dashed) line is a polynomial fit of the values obtained 
using our (the SRV) method.}
\label{fig_sai}
\end{figure}

As a first example of how our method works, we apply it to a soft-mode 
system studied by Sai, Rabe, and Vanderbilt \cite{Sai2002PRB}.
They have shown that breaking the 
inversion symmetry in perovskites by modulating their composition in a cyclic
sequence of layers produces some interesting features in their energy
landscape, apart from making them promising candidates for new materials
with useful piezoelectric and dielectric properties.
They presented results for Ba(Ti-$\delta$, Ti, Ti+$\delta$)O$_3$, where the
two species that alternate with Ti on the $B$ site are virtual atoms that
differ from Ti in that their nucleus has an defect or excess of charge equal
to $\delta$.
For different values of $\delta$ they find the $E(P)$ curves that we reproduce
in Fig.~\ref{fig_sai} (dashed lines). 
As $\delta$ increases, the double well potential becomes asymmetric (a feature
not seen in normal perovskites) until one of the minima eventually
disappears at around $\delta = 0.4$.

We have repeated their study using the same DFT methodology, but 
applying our exact method to compute $E(P)$ instead their approximate one.
We used the ABINIT \cite{abinit} code to do our calculations, with the 
Ceperley-Alder form \cite{Ceperley1980PRL} of the local-density 
approximation \cite{Kohn1965PR} (LDA) to obtain
the exchange-correlation term in DFT, a plane-wave cutoff of 35 Ha to define
the basis set, a $4 \times 4 \times 4$ Monkhorst-Pack \cite{Monkhorst1979PRB}
grid for computations in
reciprocal space, and Troullier-Martins \cite{Troullier1991PRB} norm-conserving 
pseudopotentials \cite{pseudosBaTiO3} to model the ion-electron interactions. 
The tetragonal unit cell vectors were kept fixed, with cell parameters
$a=7.547$ a.u. and $c/a=3.036$. 
Here, our results (solid lines in Fig.~\ref{fig_sai}) are very 
similar to the SRV ones; the curves only differ noticeably for large
values of the polarization.  This is not surprising, since the
dielectric behavior is dominated by the ionic response for soft-mode
systems like this one.

\begin{figure}
\center{\epsfig{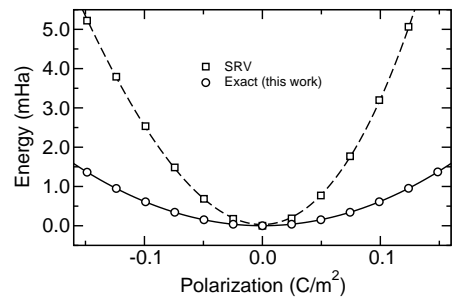}}
\caption{Energy versus magnitude of polarization in the line joining two
nearest Al and As neighbors.
The solid (dashed) line is a polynomial fit of the values obtained 
using our (the SRV) method.}
\label{fig_alas}
\end{figure}

On the other hand, the electronic response is found to be profoundly
important to the dielectric behavior
in the case of a III-V semiconductor like AlAs, as can be seen
in Fig.~\ref{fig_alas}.
Here we employed ABINIT \cite{abinit}
using the LDA \cite{Ceperley1980PRL}, a plane-wave cutoff of 9 Ha, a
$6 \times 6 \times 6$ reciprocal space grid \cite{Monkhorst1979PRB},
Troullier-Martins \cite{Troullier1991PRB} pseudopotentials \cite{pseudosAlAs},
and a theoretically optimized lattice constant of
$a=10.62$ a.u.
As can be seen in Fig.~\ref{fig_alas}, there is a drastic
difference between the $E$ versus $P$ curves computed using the SRV method
(ionic response only) and the one found with our new method (ionic and 
electronic response).
In order to quantify this difference, the dielectric constant in both cases
was computed from the curvature at the minimum of each function as
$\epsilon = 1 + 4 \pi / ({d^2E}/{dP^2})$.
When computed with the SRV method, $\epsilon=3.0$, while when we use our
new method, $\epsilon=10.3$.
The experimental result is $\epsilon=10.1$ \cite{Souza2002PRL}.

\begin{figure}
\center{\epsfig{file=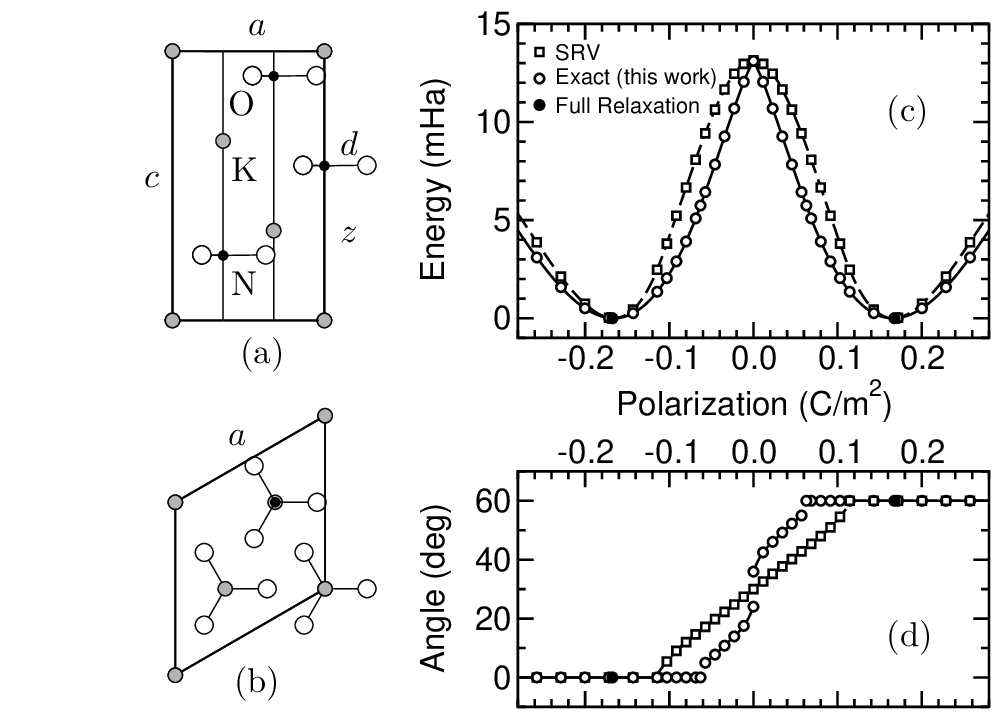,width=3.1in}}
\caption{(a) Side view of the ferroelectric phase of KNO$_3$ (phase III),
with the $c$ hexagonal lattice axis running vertically.
(b) Top view of the same structure.
(c) Energy versus polarization ${\bf P}=(0,0,P)$.
(d) Angle of rotation of the NO$_3$ group as a function of polarization $P$.}
\label{fig_kno3}
\end{figure}

Our last example involves potassium nitrate, which has an interesting
phase diagram that includes a reentrant ferroelectric phase (phase III, 
{\em R3m}), and that has been proposed as a promising material to be used
in random-access memory devices \cite{Scott1987PRB}.
The unit cell of phase III KNO$_3$ is a five-atom rhombohedral cell that
results in alternating planes of K atoms and NO$_3$ groups arranged
in such a way that the K atom is not equidistant from the NO$_3$ groups
above and below it, but instead is slightly displaced in the vertical
direction, giving rise to a polar structure.
Figs.~\ref{fig_kno3}(a)-(b) show the conventional 15-atom hexagonal cell
that is most convenient for visualization.
The calculations are performed using ABINIT \cite{abinit}, the LDA
\cite{Ceperley1980PRL}, a plane-wave cutoff of 30 Ha, a
reciprocal space grid with 6 inequivalent points, and
Troullier-Martins \cite{Troullier1991PRB} pseudopotentials \cite{pseudosKNO3}.
The theoretical structural optimization of bulk ferroelectric KNO$_3$ gives 
hexagonal lattice parameters of $a=9.68$~a.u.\ and $c=14.86$~a.u., to be 
compared with the ones found experimentally at 91 $^\circ$C, $a=10.37$~a.u.\  
and $c=17.30$~a.u.\ \cite{Nimmo1976AC}.
As for the internal degrees of freedom, the distance between a K atom and the
N atom above it is found to be $z=0.57c$ (experimentally, 
$z=0.59c$ \cite{Nimmo1976AC}), while the N--O distance is $d=2.35$ a.u.
(experimentally, $d=2.42$ a.u. \cite{Nimmo1976AC}). 
The computation of the polarization gives
0.16 C/m$^2$, higher than the experimental values of 0.08--0.11 C/m$^2$
given in Ref.~\onlinecite{Sawada1958JPSJ}.
(This disagreement is related to the reduction in volume found theoretically; 
when
we performed an optimization of the atomic positions using the experimental
lattice constants, we found a polarization value of 0.075 C/m$^2$.)

Fig.~\ref{fig_kno3}(c) shows the energy of the ferroelectric phase of KNO$_3$
as a function of the magnitude of the polarization along the $c$ axis.
As in the case of perovskites, we can see a double well potential, but 
now the change of polarization given by the movement of the NO$_3$
atoms parallel to the $c$ axis is accompanied by a rotation of 
these groups in their plane, as shown in Fig.~\ref{fig_kno3}(d).
When doing SRV calculations, both the derivative of the energy and the angle 
of rotation are continuous functions of the polarization, and the NO$_3$ group
rotates from
$0^{\circ}$ to $60^{\circ}$ as the polarization goes from negative to positive,
with the paraelectric configuration that corresponds to the maximum of $E(P)$
being the highly symmetrical one in which $z=0.5$ and the rotation angle is 
$30^{\circ}$.  However, when the new exact method is used, this behavior
changes noticeably: the $E(P)$ curve has a discontinuous derivative at $P=0$,
and the angle-rotation curve is no longer continuous.

Although it may seem puzzling that the response of the system changes
so significantly just by including the electronic response, this behavior
can be understood on the basis of a simple model in which the
energy is a function not only of polarization $P$, but also of the rotation 
angle $\theta$.  To understand the qualitative features, it is sufficient
to consider a low-order expansion
$E(P,\theta) = \cos 6 \theta + \alpha \cos 12 \theta + \beta P \cos 3 \theta 
              + P^2$.
We assume that $\beta>2 \sqrt{2}$, in which case $E(P)$ has two minima.
Then, depending on the value of $\alpha$, the behavior can be continuous
($\alpha < 1/4$) or discontinuous ($\alpha > 1/4)$ in $P$.
Here, it appears that the system was already close to this
critical value of $\alpha$, and inclusion of electronic effects
happened to shift the system from the continuous to the discontinuous regime.
A more detailed discussion of this material, and its modeling along the
lines sketched above, will be given elsewhere.  In any case, the
ability of our approach to describe the complexity of the structural
behavior of KNO$_3$ under polarization reversal provides an excellent
example of the power of the method.

To summarize, we have presented a method for finding the most
stable structural configuration of an insulating crystal when its
electric polarization is constrained to take on a given value.
Our method builds upon an earlier approach \cite{Sai2002PRB} that
makes the approximation of including only the lattice response to
applied electric fields.  By using the recently-developed
theory of finite electric fields \cite{Souza2002PRL} to include
also the electronic response of the system, we have developed a
new approach that is, instead, exact.  We have illustrated the
method by obtaining $E(P)$ curves for three rather different kinds
of insulating materials, and have illustrated how the method is
capable of describing the complexity of the nonlinear structural
and dielectric response.

\begin{acknowledgments}
O.~D. acknowledges useful discussions with Jos\'e-Luis Mozos, and the
instruction and encouragement received from him during visits to the
ICMAB in Barcelona.
The authors thank J.~Scott for suggesting the application to KNO$_3$.
This work was supported by ONR Grants N0014-05-1-0054, N00014-00-1-0261,
and N00014-01-1-0365.
\end{acknowledgments}



\end{document}